\documentclass[pra,aps,superscriptaddress,showpacs,floatfix,tightenlines,twocolumn]{revtex4-1}
\usepackage{graphicx}
\usepackage{dcolumn}
\usepackage{bm}
\usepackage{amssymb}
\usepackage{amsmath}
\usepackage{url}
\usepackage{amsthm}
\usepackage{layout}
\usepackage{epsfig}
\usepackage{graphicx}
\usepackage{epstopdf}
\usepackage{float}
\usepackage{subfigure}
\usepackage[hyperindex,breaklinks]{hyperref}
\usepackage{booktabs}
\usepackage{amsmath}
\usepackage{mathrsfs}
\usepackage{tabularx}
\usepackage{algorithm}
\usepackage{algorithmic}
\usepackage{xcolor}

\newcommand{\ket}[1]{|#1\rangle}
\newcommand{\bra}[1]{\langle #1 |}

\def\Tr{{\rm Tr}}
\newtheorem{theorem}{Theorem}
\newtheorem{lemma}{Lemma}

\usepackage{enumitem}
\usepackage{amsmath}
\usepackage{amssymb}
\setlist[itemize]{leftmargin=*,labelindent=0pt}

\begin{document}

\title{Reducing  Complexity of Shadow Process Tomography with Generalized Measurements}
\author{Haigang Wang}
\affiliation{College of Mathematics, Taiyuan University of Technology, Taiyuan, 030024, China}


\author{Kan He}
\email{hekan@tyut.edu.cn}
\affiliation{College of Mathematics, Taiyuan University of Technology, Taiyuan, 030024, China}

\date{\today }
\begin{abstract}
Quantum process tomography (QPT) is crucial for advancing quantum technologies, including quantum computers, quantum networks and quantum sensors.
Shadow process tomography (SPT) utilizes the Choi isomorphism to map QPT to shadow state tomography (SST), significantly reducing the sample complexity for extracting information from quantum processes. However, SPT relies on random unitary operators and complicates the determination of the optimal unitary operator that minimizes the shadow norm, which is the key factor influencing the sample complexity. In this work, we propose a generalized SPT framework that minimizes the shadow norm by replacing unitary operators with generalized measurements (POVMs). This approach, termed shadow process tomography with POVMs (POVM-SPT), uses convex optimization to identify the optimal POVM for minimizing the shadow norm, thereby further reducing sample complexity. We demonstrate the identification of the optimal POVM through numerical simulations and provide the corresponding optimization algorithms. 
Our numerical experiments demonstrate that POVM-SPT achieves a substantial reduction in shadow norm compared to conventional SPT, with an approximate 7-fold improvement for single-qubit input states and a remarkable $2^{180}$-fold enhancement for 64-qubit input states. These results reveal that POVM-SPT offers significant advantages in simplifying SPT tasks, particularly for large-scale quantum systems.

\end{abstract}

\maketitle


\section{Introduction}

\begin{table*}[t]
    \centering
     \caption{
     Comparison of squared shadow norms between $ \|d \cdot \rho^T \otimes X \|^2_{P^{\rm tot}}$ (for SPT) and $ \|d \cdot \rho^T \otimes X \|^2_{E^{\rm tot}}$ (for POVM-SPT). For details of squared shadow norms, see Eq. (\ref{le1}) and (\ref{lel-p}).
     The first row of data corresponds to the case where $\rho = |0\rangle\langle0|, X = \sigma_x$ (a Pauli matrix). The second row of data corresponds to the case where $\rho=\frac{(|0\rangle+|1\rangle)( \langle 0| +\langle 1|)}{2},X=\sigma_y$ (a Pauli matrix).
     The first two columns present results for Pauli unitary measurements (PUMs) and Clifford unitary measurements (CUMs), respectively, while the last three columns correspond to POVMs with $N = 4, 6$ and $ 8$ effects.
     }
    \label{tab:example1}
    \begin{tabularx}{0.99\textwidth}{cccccccc}
        \hline
        \hline
        \textbf{$(\rho, \, X)$} & \quad \quad PUM \,\,\,\,  &\quad \quad CUM \,\, &\quad \quad $\rm POVM_{ N=4}$ \,\,\,\, &\quad \quad$\rm POVM_{ N=6}$ \,\,\,\, &\quad \quad $\rm POVM_{ N=8}$ \,\  \\
        \hline
        \textbf{$\rho=|0\rangle\langle0|, X=\sigma_x$}  &\quad \quad 64 \,\,&\quad \quad 448 \,\, & \quad \quad4.12\,\, &\quad \quad 4.07 \,\, &\quad \quad 4.06 \,\  \\
        \textbf{$\rho=\frac{(|0\rangle+|1\rangle)( \langle 0| +\langle 1|)}{2},X=\sigma_y$}  &\quad \quad 64 \,\,&\quad \quad 480 \,\,\, &\quad \quad 4.08 \,\, &\quad \quad 4.07 \,\, &\quad \quad 4.07 \,\  \\
        \hline
        \hline
    \end{tabularx}
\end{table*}

In quantum mechanics, characterizing dynamical processes is a foundational problem. A quantum process, typically described  by a quantum channel, plays a central role in various areas of quantum technology. For instance, in quantum networks \cite{QN1,QN2}, quantum processes model the generation of entanglement between nodes \cite{r1}; meanwhile, in quantum computing, they serve as the target for implementing quantum circuits \cite{quantum.c}. Beyond their technological significance, the study of quantum channels provides critical insights into fundamental questions across multiple domains, including the investigation of entanglement in the context of gravity \cite{r5, r7}.

Quantum process tomography (QPT) is the task of fully characterizing an unknown quantum process from measurement data, and has become a powerful tool in quantum technology \cite{r8}. Early QPT \cite{q7,q8} employed linear inversion techniques to reconstruct quantum channels using informationally complete datasets, which were obtained by feeding known input states into the quantum process and performing full quantum state tomography (QST) \cite{qst1,qst2} on the resulting outputs. The channel-state duality introduced by Choi and Jamiołkowski enabled a more efficient approach: applying the channel to a maximally entangled state between the system and an ancilla, followed by QST to directly reconstruct the channel's Choi matrix \cite{4,5,q12}. Further statistical techniques, such as maximum likelihood estimation \cite{q13,q15} and Bayesian inference \cite{q17,q18}, have been adapted for QPT.

However, conventional QPT methodologies require an extensive set of measurements. Specifically, the number of measurements scales polynomially with the Hilbert space dimension and exponentially with the number of qubits \cite{rr8}. In many applications, full characterization of a quantum channel is unnecessary. 
Often, only specific information about the quantum channel is of interest, such as the expectation values of the resulting states with respect to certain observables, the unitary properties of the channel, the bit-flip conversion probability, and so on \cite{rr9, 3, prx-ml}.
To address this, shadow process tomography (SPT) was proposed \cite{3,prr-spt}, transforming the problem into shadow state tomography (SST) \cite{ref3, povm-shadow} via the Choi isomorphism. SPT enables the extraction of essential information about quantum processes with fewer samples than QPT. 
Specifically, SPT/SST involves four steps:  
(1) selecting a random unitary operator and applying it to the Choi state,  
(2) performing an ideal projection measurement,  
(3) applying the inverse unitary operator to the post-measurement state, 
(4) using a reversible linear map to obtain a classical shadow \cite{ref3, 1}.
Repeated application of this procedure generates a series of shadows, allowing predictions of key properties—such as expectation values, entanglement entropy, and fidelity—--with a relatively small sample size \cite{ref3}. Nevertheless, the sample complexity (or shadow norm) of SPT/SST critically depends on the choice of random unitary operators. Although some studies aim to reduce the upper bound on the shadow norm \cite{opt,opt2,opt3}, determining the optimal unitary operator that minimizes the shadow norm remains a significant challenge.

In this paper, inspired by the work in \cite{0}, we propose a generalized SPT to reduce the sample complexity: rather than relying on random unitary operators and projection measurements, we utilize least-square estimator \cite{ref1,ref2,0,r8} and generalized measurements (POVMs) to perform SST for Choi states. We refer to this method as shadow process tomography with POVMs (POVM-SPT). 
POVM-SPT constructs classical shadows through least-square estimators on the measurement outcomes of the Choi states. This approach fundamentally differs from methods relying on post-measurement states \cite{3, prr-spt}. The SPT based on projection measurements in \cite{3} is a special case of POVM-SPT, as the set of POVMs encompasses projection measurements.
By employing informationally complete POVMs, POVM-SPT eliminates the need for a large number of unitary operators. 
Most importantly, POVM-SPT enables convex optimization over the set of POVMs to identify the optimal POVM that minimizes the shadow norm.
Table~\ref{tab:example1} presents a simple single-qubit case study comparing the optimal squared shadow norms achieved by our method with the theoretical upper bounds from~\cite{3}, clearly demonstrating substantial improvements in the shadow norm reduction.
Furthermore, our numerical experiments establish that for 64-qubit input states, our method outperforms conventional SPT by a remarkable factor of $2^{180}$ in sample complexity reduction. This dramatic improvement provides crucial theoretical foundations for implementing large-scale QPT.

The paper is structured as follows: In Sec.~\ref{sec2}, we provide a brief review to the SST. In Sec.~\ref{sec3}, we derive the POVM-SPT and the sample complexity. In Sec.~\ref{sec4}, we first prove that the shadow norm obtained from POVM-SPT is always less than or equal to the shadow norm obtained from traditional SPT. Then, we provide a detailed explanation of how POVM-SPT minimizes the shadow norm through the simulated annealing algorithm. Finally, in Sec.~\ref{sec5}, we summarize our findings and outline potential future research directions

\section{Shadow state tomography with generalized measurements}\label{sec2}
In this section, we briefly summarize SST with generalized measurements, which is proposed  by Nguyen {\it et al.} in \cite{0}.

A generalized measurement, known as a positive operator-valued measure (POVM), is represented by a collection of positive operators, termed effects, denoted as \( E = \{E_1, E_2, \ldots, E_N\} \). These effects are complete, satisfying the condition \( \sum_{k=1}^{N} E_k = \mathbb{I} \). Each POVM \( E \) associates a quantum state \( \rho \) with a probability distribution corresponding to the measurement outcomes, thus establishing the mapping  \( \Phi_E \) defined as:
\begin{equation}\label{1}
\Phi_E(\rho) = \{ \text{Tr}(\rho E_k) \}_{k=1}^{N}.
\end{equation}
When the quantum state $\rho$ is copied $M$ times and  POVMs are repeatedly performed on each copy, it generates a sequence of outcomes \( \{k_i\}_{i=1}^M \), where \( k_i \in \{1, \ldots, N\} \) indicates the result of the \( i \)-th measurement. 
The frequency of the measurement outcome \( k \) is denoted as \( p_k \), which serves as an approximation for \( \Tr (\rho E_k) \). Consequently, a frequency vector \( \vec{p} \in \mathbb{R}^N \) is obtained, which can be used for reconstructing the density operator \( \rho \). According to the least-square estimator \cite{ref1,ref2}, the estimated state is obtained by:
\begin{equation}\label{3}
\chi_{\text{LS}}(\vec{p}) = \arg \min_{\tau:states } \sum_{k=1}^{N} \left[ \text{Tr}(\tau E_k) - p_k \right]^2.
\end{equation}
The effects \( \{E_k\}_{k=1}^N \) are assumed to span the entire operator space, thereby rendering the  POVM \( E \) informationally complete. This implies that \( \Phi_E^\dagger \Phi_E \) is invertible, allowing the solution of the least-square estimator \( \chi_{\text{LS}}(\vec{p}) \) to be expressed as:
\begin{equation}\label{8}
\chi_{\text{LS}} = (\Phi_E^{\dagger} \Phi_E)^{-1} \Phi_E^{\dagger}.
\end{equation}

SST starts by associating each  outcome \( k \) with the distribution \( \vec{q}_k = \{\delta_{k,l}\}_{l=1}^{N} \), where \( \delta_{k,l} = 1 \) if \( l = k \) and \( 0 \) otherwise. This single measurement outcome can be utilized to obtain a noisy estimate of the density operator \( \rho \), referred to as the classical shadow \cite{1,ref3,0}, expressed as:
\begin{equation}\label{9}
\hat{\rho}_k = \chi_{\text{LS}}(\vec{q}_k).
\end{equation}
Crucially, the adjoint map $\Phi_E^\dagger$ satisfies  \( \Phi_E^\dagger(\vec{q}_k) = E_k \). By defining the operator \( C_E = \Phi_E^\dagger \Phi_E \),  we derive the key identity 
$C_E(\rho) = \sum_{k=1}^{N} \operatorname{Tr}(\rho E_k) E_k.$
Consequently, the classical shadow admits an equivalent representation in terms of   $C_E^{-1}$:
\begin{equation}\label{shadow}
\hat{\rho}_k = C_E^{-1}(E_k).
\end{equation}
If we fix a measurement, then $\hat{\rho}_k$ will be uniquely determined (for example, see Appendix \ref{appendix A}).
In the case of an infinite number of measurements, the statistical average of classical shadows converges surely to the true density operator:
\begin{equation}\label{13}
\rho = \lim_{M \to \infty} \frac{1}{M} \sum_{i=1}^{M} \hat{\rho}_{k_i}.
\end{equation}

Each classical shadow, as defined in Eq. (\ref{shadow}), acts as an intermediate processed data for the subsequent computation of observables. For a given observable \( X \), each \( \hat{\rho}_k \) generates an estimator $\hat{x}_k = \text{Tr}(\hat{\rho}_k X)$ for \( \langle X \rangle \).
Utilizing a dataset \( \{k_i\}_{i=1}^M \), the sample average:
$\frac{1}{M} \sum_{i=1}^M \hat{x}_{k_i}$
converges to the expectation of $X$ when $M \to \infty$.
 The asymptotic convergence rate of this estimation is governed by the estimator’s variance:
\begin{equation}\label{14}
\text{Var}(\hat{x}_k) = \sum_{k=1}^N \text{Tr}(\hat{\rho}_k X)^2 \text{Tr}(\rho E_k) - \langle X \rangle^2.
\end{equation}
By neglecting the second term, one obtains an upper bound for the variance, leading to the definition of the shadow norm of $X$ \cite{ref3,0}:
\begin{equation}\label{15}
\text{Var}(\hat{x}_k) \leq \| X \|^2_E = \lambda_{\text{max}} \left( \sum_{k=1}^N \text{Tr}(\hat{\rho}_k X)^2 E_k \right),
\end{equation}
where $\lambda_{\text{max}}(\cdot)$
  denotes the largest eigenvalue of the corresponding operator. For a family of observables $\mathcal{X}$, the worst-case variance is characterized by:
\begin{equation}\label{16}
\kappa^2_E(X) = \max \{ \| X \|^2_E : X \in \mathcal{X} \},
\end{equation}
with smaller $\kappa^2_E(X)$ indicating higher estimation accuracy \cite{0}.

SST is scalable for $n$-qubit systems. Let each qubit undergo a local POVM    $E^{(i)} = \{ E^{(i)}_{k^{(i)}} \}_{k^{(i)}=1}^{N_i} \, \, (      i  =  1,   \dots  ,  n)$, which collectively defines a global POVM    $E^{tot}$   with effects
$ E^{\text{tot}}_ k =  \bigotimes_{i=1}^nE^{(i)}_{k^{(i)}}$.
The corresponding global classical shadow factorizes as $\hat{\rho}^{\text{tot}}_k = \bigotimes_{i=1}^n \hat{\rho}^{(i)}_{k^{(i)}}$ , 
where \( \hat{\rho}^{(i)}_{k^{(i)}} \) is the classical shadow corresponding to the measurement \( E^{(i)} \) on the \( i \)-th qubit. For observables decomposing tensorially $ X = \bigotimes_{i=1}^n X^{(i)}$,
the shadow norm factorizes multiplicatively:
\begin{equation} \| X \|_{E^{\rm tot}} = \| X^{(1)} \|_{E^{(1)}} \| X^{(2)} \|_{E^{(2)}} \cdots \| X^{(n)} \|_{E^{(n)}}. \end{equation}

\section{ Shadow process tomography with generalized measurements}\label{sec3}
Just as SST predicts the expectations of observables for a set of unknown quantum states, the objective of SPT is to predict the expectation values  of observables for a set of known quantum states passing through an unknown channel \cite{3}. Given a set of known states $\{\rho_l\}_{l=1}^G$, and a quantum channel $\mathcal{E}: \mathbb{C}^{d \times d} \to \mathbb{C}^{d \times d}$ which is completely positive and trace-preserving, the states are transformed into $\{\mathcal{E}(\rho_l)\}_{l=1}^G$. 
Our aim is to predict important information about the quantum channel, such as $\{ \text{Tr}[\mathcal{E}(\rho_l) X_j] \}_{l,j=1}^{G,H}$ for a set of observables $\{X_j\}_{j=1}^H$, using as few measurements as possible.
\subsection{Choi isomorphism}
\begin{figure}
    \centering
    \includegraphics[width=1.0\linewidth]{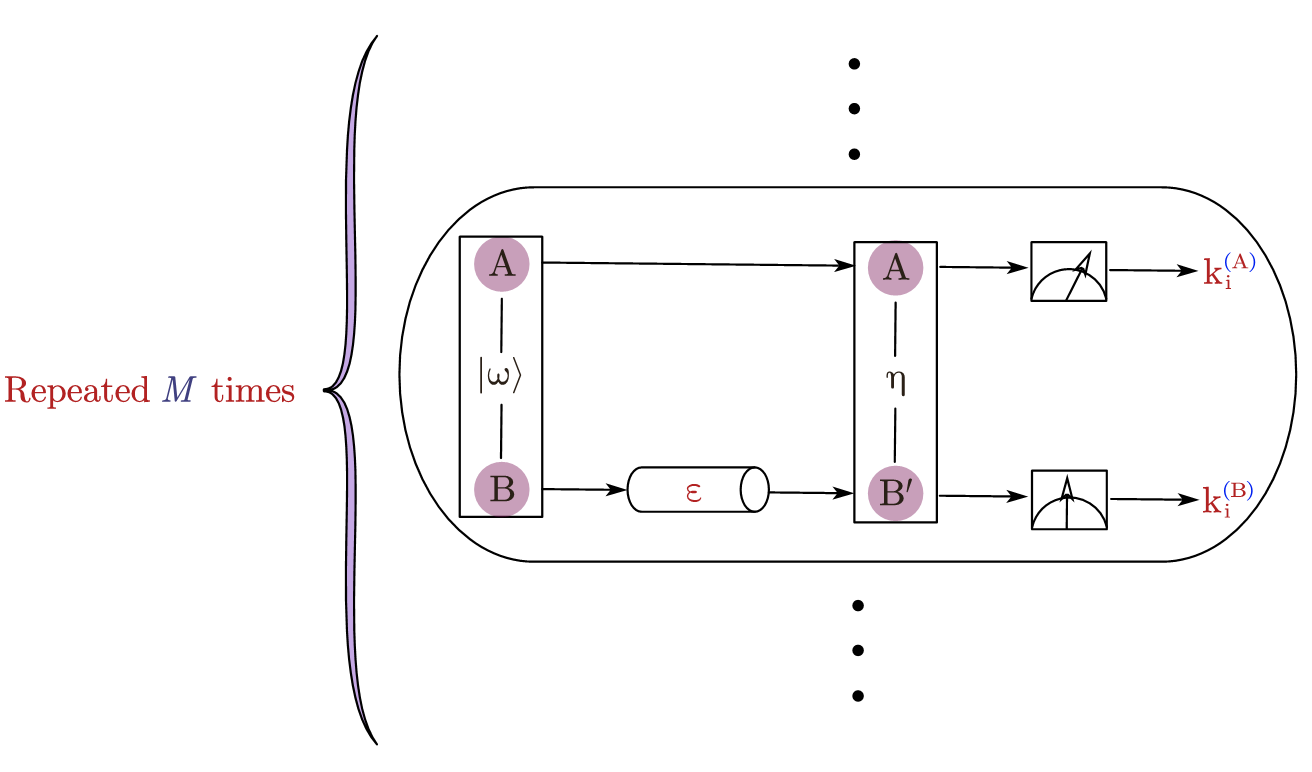}
   \caption{By using the Choi isomorphism, the task of measuring $\mathcal{E}(\rho)$ is transformed into measuring the Choi state. Quantum systems A and B share a pair of maximally entangled particles. The particle from system B passes through an unknown channel $\mathcal{E}$, while the particle from system A goes through an identity channel $\mathbb{I}$, resulting in the Choi state $\eta$. The $i$th POVM measurement is performed on systems A and B, yielding corresponding measurement results $k^{(A)}_i$ and $k^{(B)}_i$. Repeating this process $M$ times allows us to obtain the frequency corresponding to each measurement results.}
    \label{choi}
\end{figure}

To apply the SST to a quantum process $\mathcal{E}$, we employ the  Choi-Jamiolkowski isomorphism, which maps a process into a density matrix \cite{3,4,5}. Consider the unnormalized maximally entangled state in the tensor product space $\mathbb{H}^{\rm A} \otimes \mathbb{H}^{\rm B}$, where $\mathbb{H}^{\rm A}$ and $\mathbb{H}^{\rm B}$ are $d$-dimension Hilbert spaces:
\begin{equation}\label{17}
|\omega\rangle = \sum_{n=0}^{d-1} |n\rangle_{\rm A} \otimes |n\rangle_{\rm B}.
\end{equation}
The Choi state is then constructed as:
\begin{equation}\label{18}
\eta = (\mathbb{I}_{\rm A} \otimes \mathcal{E}_{\rm B})(|\omega\rangle\langle \omega|),
\end{equation}
where    $\mathbb{I}_{\rm A}$   is the identity operator on $\mathbb{H}^{\rm A}$.
Notably,  \( \eta \) is proportional to a density operator with normalization factor \( d^{-1} \).
For a Choi state $\eta$ of size $d^2 \times d^2$ and an input state  $\rho$ of size $d \times d$, the result of acting $\mathcal{E}$ to $\rho$ can be computed as:
\begin{equation}\label{19}
\mathcal{E}(\rho) = \operatorname{Tr}_{\rm A}[(\rho^T \otimes \mathbb{I}_{\rm B}) \eta].
\end{equation}
Furthermore, given an observable \( X  \) of size $d\times d$, we have the relationship:
\begin{equation}
\operatorname{Tr}[\mathcal{E}(\rho) X] = \operatorname{Tr}[\eta (\rho^T \otimes X)].
\end{equation}
By treating  \(\rho^T \otimes X\) as a generalized observable acting on $\eta$, 
the problem of estimating $\text{Tr}[\mathcal{E}(\rho) X]$ reduces to shadow tomography of the Choi state $\eta$ .

For example, when $\mathcal{E}$ is single qubit depolarizing channel, $\mathcal{E}(\rho)$ is a two-qubit Werner state. As shown in Fig. \ref{choi}, we can perform POVM measurements on it to obtain a series of measurement results \cite{ieee, qmqm}.
These measurement results can be further used to generate the classical shadow of the Choi state.
Although executing measurements on the Choi states requires a significant consumption of highly entangled states, we do not need to prepare the initial states $\{\rho_l\}_{l=1}^G$ repeatedly or apply the channel $\mathcal{E}$ repeatedly.  This establishes a formal equivalence between the SPT and the SST framework applied to \(\eta\). Therefore, we only need to consider how to perform SST on the Choi state $\eta$.

Note that our goal is not merely to predict the expectation values of the output states with respect to observables (which could have been done directly on the output state via SST). This would result in the loss of some important properties of the quantum channel. Instead, the Choi state \(\eta\) is an equivalent characterization of the quantum channel that contains all the information about the quantum channel. By adjusting the forms of $X$ and $\rho$, we can predict the expectation value of the Choi state \(\eta\) with respect to different observables $\rho^T \otimes X$, thereby predicting some important properties of the quantum channel, such as transition probabilities, multitime correlation functions, unitarity verification, and so on \cite{3}.

\subsection{Constructing classical shadows for Choi states using POVMs}

 We consider a quantum system prepared in the normalized Choi state \(\frac{\eta}{d}\) and perform the POVM \(E^{\rm tot}\).
 Let \(E^{(1)} = \{ E^{(1)}_{k^{(1)}} \}_{k^{(1)}=1}^{N}\) and \(E^{(2)} = \{ E^{(2)}_{k^{(2)}} \}_{k^{(2)}=1}^{N}\) denote POVMs acting on the Hilbert spaces \(\mathbb{H}^{\rm A}\) and \(\mathbb{H}^{\rm B}\), respectively.
The effect of combined POVM on the $\mathbb{H}^{\rm A} \otimes \mathbb{H}^{\rm B}$ are defined as \(E^{\rm tot}_k = E^{(1)}_{k^{(1)}} \otimes E^{(2)}_{k^{(2)}}\),
forming the global POVM  \(E^{\rm tot} = \{ E^{\rm tot}_k \}_{k=1}^{N^2}\).
Consequently, the classical shadow associated with the effect \(E^{\rm tot}_k\) on the 
 $\mathbb{H}^{\rm A} \otimes \mathbb{H}^{\rm B}$  is given by:
\begin{equation}\label{choi-shadow}
    \frac{\hat{\eta}_k}{d} = \hat{\rho}^{(1)}_{k^{(1)}} \otimes \hat{\rho}^{(2)}_{k^{(2)}},
\end{equation}
where \(\hat{\rho}^{(i)}_{k^{(i)}}\) represents the classical shadow from measurement \(E^{(i)}\) on the \(i\)-th subsystem.
By repeating \(E^{\rm tot}\) on  $M$ copies of \(\frac{\eta}{d}\), we obtain outcomes \(\{ k_i \}_{i=1}^M\) and construct $M$  shadows \(\frac{\hat{\eta}_1}{d}, \ldots, \frac{\hat{\eta}_M}{d}\). The  empirical average of these shadows \(\frac{1}{M} \sum_{i=1}^{M} \frac{\hat{\eta}_{k_i}}{d}\) provides a consistent estimator for the normalized Choi state, satisfying:
\begin{equation}
 \frac{\eta}{d} = \lim_{M \to \infty} \frac{1}{M} \sum_{i=1}^{M} \frac{\hat{\eta}_{k_i}}{d}.
\end{equation}

 Requiring the estimator \( \hat{\eta} \) to closely approximate the Choi state \( \eta \) in operator norm is often prohibitively restrictive \cite{3}. Instead, we adopt a relaxed fidelity criterion: the estimator must reproduce the expectation values of target observables with high precision.
Hence, we define the estimator for the expectation \(\langle X \rangle = \operatorname{Tr}[\mathcal{E}(\rho) X]\) as:
\begin{equation}
\hat{x}_{k_i} = \operatorname{Tr}[\hat{\eta}_{k_i} (\rho^T \otimes X)],    
\end{equation}
where \(\hat{x}_{k_i}\) corresponds to the \(i\)-th measurement outcome. 
Collecting  \(M\) outcomes \(\{k_i\}_{i=1}^M\), 
we compute $M$ independent estimators \(\hat{x}_{k_1}, \hat{x}_{k_2}, \ldots, \hat{x}_{k_M}\).
As $M \to \infty$, the average of these  estimators converges to  \(\langle X \rangle\):
\begin{equation}\label{13aa}
 \langle X \rangle = \lim_{M \to \infty} \frac{1}{M} \sum_{i=1}^M \hat{x}_{k_i}.
\end{equation}
The asymptotic convergence rate of this estimator  is related to the variance.

\begin{lemma}[Variance Bound]\label{lemma 1}
For a fixed quantum channel $\mathcal{E}: \mathbb{C}^{d \times d} \to \mathbb{C}^{d \times d}$, POVM $E^{\rm tot} = \{ E^{\rm tot}_k \}_{k=1}^{N^2}$, observable $X \in \mathbb{C}^{d \times d}$, and known state $\rho \in \mathbb{C}^{d \times d}$, define the single-shot estimator:
\[
\hat{x}_{k} = \operatorname{Tr}\left[\hat{\eta}_k (\rho^T \otimes X)\right], \quad \hat{\eta}_k = d \cdot \hat{\rho}^{(1)}_{k^{(1)}} \otimes \hat{\rho}^{(2)}_{k^{(2)}}.
\]
The estimator's variance satisfies:
\begin{equation}\label{le1}
\begin{aligned}
\operatorname{Var}(\hat{x}_k) 
&\leq \left\| d \cdot \rho^T \otimes X \right\|^2_{E^{\rm tot}} \\
&= \lambda_{\rm max}\left( \sum_{k=1}^{N^2} \operatorname{Tr}[\hat{\eta}_k (\rho^T \otimes X)]^2 E^{\rm tot}_k \right),
\end{aligned}
\end{equation}
where $\lambda_{\rm max}(\cdot)$ is the spectral norm. (Proof: Appendix~\ref{appendix lem}.)
\end{lemma}

Lemma~\ref{lemma 1} establishes the shadow norm $\| \cdot \|_{E^{\rm tot}}$ as the key metric governing estimation variance in POVM-SPT. Crucially, smaller shadow norms enable higher precision through variance suppression.

The framework naturally extends to estimating $HG$ expectation values $\{\operatorname{Tr}[\mathcal{E}(\rho_l) X_j]\}_{l,j=1}^{G,H}$. For robustness against statistical outliers, we employ the median-of-means estimator \cite{6}:
\begin{equation}\label{20}
\begin{aligned}
&\hat{x}_j(\rho_l, M/K, K) \\& = \mathrm{Median} \Bigg\{ \hat{x}^{(1)}_j(\rho_l,M/K,1),\cdots,\hat{x}^{(k)}_j(\rho_l,M/K,1),  \\& \quad  \cdots \hat{x}^{(K)}_j(\rho_l,M/K,1) \Bigg\},
\end{aligned}
\end{equation}
where
\begin{equation}\label{21}
\hat{x}^{(k)}_j(\rho_l, M/K, 1) = \frac{K}{M} \sum_{i=(k-1)M/K +1}^{kM/K} \operatorname{Tr}[\hat{\eta}_{k_i} (\rho_l^T \otimes X_j)].
\end{equation}
This approach achieves exponential concentration in estimation error with respect to the number of batches $K$, significantly outperforming sample-mean estimators \cite{ref3,3}.

\begin{theorem} [The sample complexity] \label{theorem 1}
Let \( \{X_j\}_{j=1}^H \) be observables and \( \{\rho_l\}_{l=1}^G \) be known quantum states. For any \(\epsilon >0\), \(\delta >0\), and POVM \( E^{\mathrm{tot}} = \{E^{\mathrm{tot}}_k\}_{k=1}^{N^2} \), define:
\begin{align}
K &= 2\ln(2HG/\delta), \label{eq:K} \\
\frac{M}{K} &= \frac{34}{\epsilon^2} \max_{\substack{1\leq j \leq H \\ 1\leq l \leq G}} \| d \cdot \rho_l^T \otimes X_j \|^2_{E^{\mathrm{tot}}} \label{eq:M}
\end{align},
where \(\| \cdot \|_{E^{\mathrm{tot}}}\) denotes the shadow norm in Eq. (\ref{le1}). Then \(M\) independent shadows enable accurate prediction via median-of-means estimation:
\begin{equation}
\left| \hat{x}_j(\rho_l,M/K,K) - \operatorname{Tr}[\mathcal{E}(\rho_l)X_j] \right| \leq \epsilon ,\quad   \label{eq:error}
\end{equation}
$ \forall \,\,\, 1\leq j\leq H, 1 \leq l \leq G$,  with probability \(\geq 1-\delta\). (Proof: Appendix \ref{appendix th1}.)
\end{theorem}

The sample complexity \(M\) satisfies:
\begin{equation}\label{eq:sample}
\begin{aligned}
M = \frac{68\ln(2HG/\delta)}{\epsilon^2} \kappa^2_{E^{\mathrm{tot}}} 
= \mathcal{O}\left(\frac{\ln(HG) \cdot \kappa^2_{E^{\mathrm{tot}}}}{ \epsilon^{2}} \right)
\end{aligned}
\end{equation}
for predicting \(HG\) functionals \(\operatorname{Tr}[\mathcal{E}(\rho_l)X_j]\). Here the key parameter:
\begin{equation}\label{eq:kappa}
\kappa^2_{E^{\mathrm{tot}}} := \max_{\substack{1\leq j \leq H \\ 1\leq l \leq G}} \| d \cdot \rho_l^T \otimes X_j \|^2_{E^{\mathrm{tot}}}
\end{equation}
characterizes the maximal squared shadow norm over the composite observable set \(\mathcal{X} = \{d \cdot \rho_l^T \otimes X_j\}_{j,l=1}^{H,G}\).

Now, let us summarize the step-by-step protocol of POVM-SPT for the estimation of a set of observables \( \{X_j\}_{j=1}^H \) :

1. Given a set of known state \( \{\rho_l\}_{l=1}^G \) and an unknown quantum channel $\mathcal{E}$.

2. The Choi state $\eta = (\mathbb{I} \otimes \mathcal{E}_B)(|\omega\rangle\langle \omega|)$ is prepared for investigation and the measurement \( E^{\rm tot} = \{E^{\rm tot}_k\}_{k=1}^{N^2} \) is carried out. This is repeated  \( M \) times, and the string of outcomes \( \{ k_i \}_{i=1}^M \) is recorded.

3. According to the outcomes, classical shadows \( \{ \hat{\eta}_{k_i} \}_{i=1}^M \) for Choi states are computed using formulas given in Eq. (\ref{choi-shadow}).

4. The mean values of \( \{X_j\}_{j=1}^H \) for \( \{\mathcal{E}(\rho_l)\}_{l=1}^G \) are estimated by
the median-of-means estimator $\hat{x}_j(\rho_l,M/K,K)$ given in Eq. (\ref{20}) and Eq. (\ref{21}).

\section{Optimizing  shadow process tomography}\label{sec4}

Although we have derived the sample complexity of POVM-SPT, it is contingent upon the selection of measurements, and different measurement choices can lead to varying sample complexities. Therefore, determining how to choose measurements to minimize sample complexity is a key issue that we need to address.
That is to find the optimal  POVM \( {E^{\rm tot}}^* \) that minimizes the maximal shadow norm:
\begin{equation}\label{kappa_min}
{E^{\rm tot}}^* = \arg\min_{E^{\rm tot}} \kappa^2_{E^{\rm tot}}(\mathcal{X}).
\end{equation}

Here, we highlight that the case presented in \cite{3} serves as a special example within our study. According to the definition of the shadow norm in \cite{3}, we can derive the following (for details, see Appendix \ref{appendix b})

\begin{equation}\label{lel-p}
    \begin{aligned}
        & \|d \cdot \rho^T \otimes X \|^2_{P^{\rm tot}} \\&= \lambda_{\text{max}} \left\{ \sum_{k=1}^{N^2} \text{Tr}[\hat{\eta}_k (\rho^T \otimes X)]^2  P^{{\rm tot}}_k \right\},
    \end{aligned}
\end{equation}
where \( P^{\rm tot} = \{P^{{\rm tot}}_k\}_k^{N^2} \) represents a random Clifford measurement (CUM) or random Pauli measurement (PUM), both of which are special types of projection measurements \cite{ref3}.

\begin{figure}[t]
  \centering
  \includegraphics[width=0.9\linewidth]{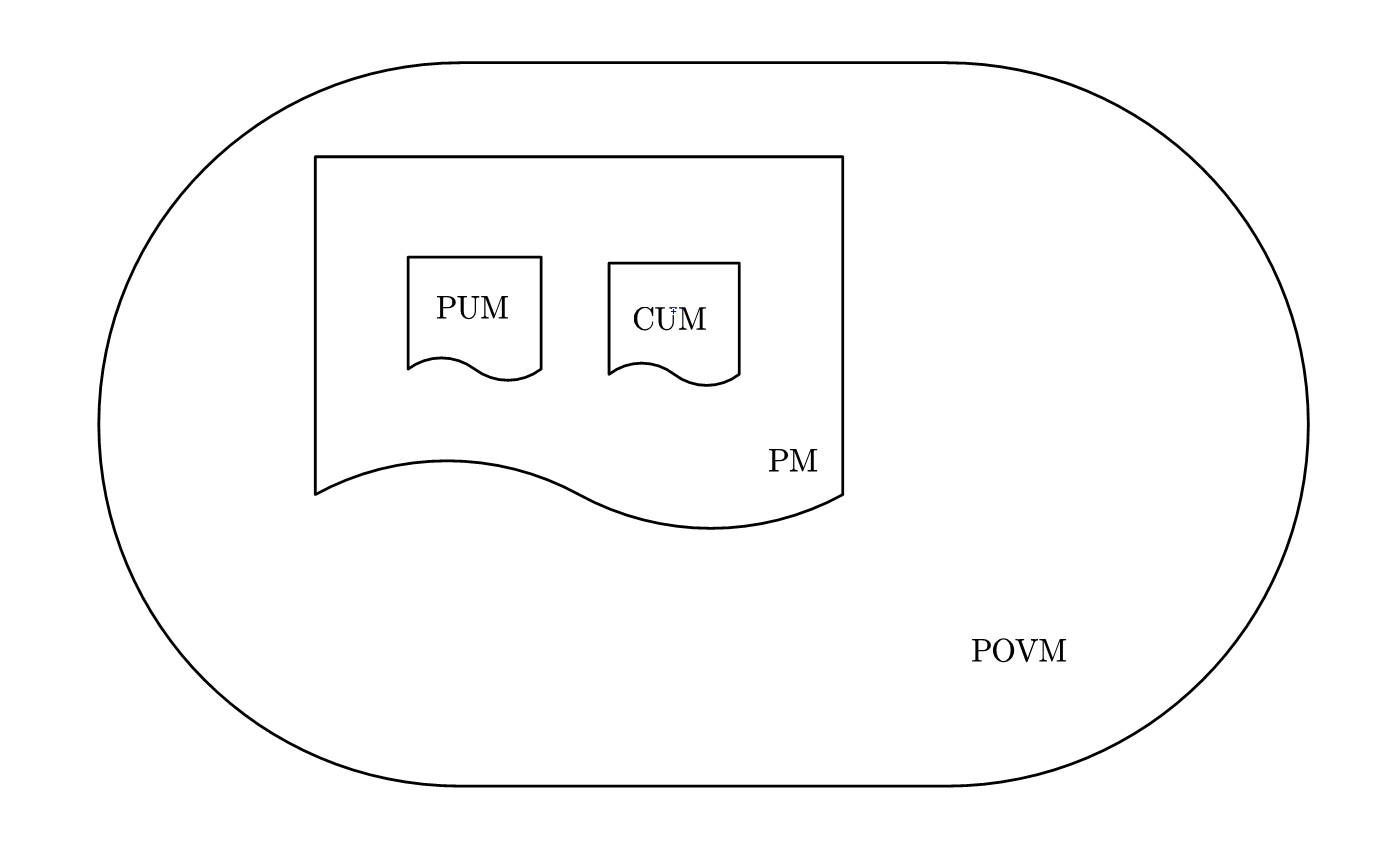}\\
  \caption{Inclusion hierarchy of measurement types: POVM, projection measurement (PM), Pauli unitary measurement (PUM) and Clifford unitary measurement (CUM).}\label{measurement}
\end{figure}

\begin{theorem}\label{theorem 2}
   The optimal maximum squared shadow norm obtained in the POVM set is always less than or equal to that obtained in the projection measurement set: 
   \(\kappa^2_{E^{\rm tot^*}} \leq \kappa^2_{P^{\rm tot^*}}.\) (Proof: Appendix \ref{appendix th2}.)
\end{theorem}

As demonstrated in \cite{3}, the optimization restricted to randomized unitaries is computationally intractable even for SST. However, when extended to all POVMs, this becomes an optimization problem over a convex domain that includes the PUMs and CUMs.

\begin{subsection}{Single qubit systems}
We start by studying the case of  single qubit systems. Although POVM-SPT tends to serve many-body systems, the case of  single qubit is fundamental and can be naturally extended to many-body systems.

To represent any POVM in single qubit system, we use the Pauli matrices $\{ \sigma_i \}_{i=1}^4 = \{ I, \sigma_x, \sigma_y, \sigma_z \}$ as a basis for the operator space. Then, any operator $X$ on the qubit system can be expressed as a real vector with four components:$
X = \frac{1}{2} \sum_{i=1}^4 x_i \sigma_i$,
where \( x_i = \text{Tr}(\rho \sigma_i) \). For operators \( X \) and \( Y \) represented by vectors \( \vec{x} \) and \( \vec{y} \) , respectively, the Hilbert-Schmidt inner product simplifies to:
$\text{Tr}(XY) = \frac{1}{2} \sum_{i=1}^4 x_i y_i$. 
Notably, the identity operator $\mathbb{I}_2$ corresponds to the Bloch vector \( (2, 0, 0, 0)^T \).

In this Bloch representation, each effect $E_k$ adopts the unified representation:
\begin{equation}
E_k=\frac{2}{N} \begin{pmatrix} 1  \\ \vec{r}_k  \end{pmatrix}, 
\end{equation}
where \( \|\vec{r}_k\| \leq 1 \) and the condition $\sum_{k=1}^N \vec{r}_k = 0$ holds.
Hence, the effects of a POVM \( E^{\rm tot} = \{E^{\rm tot}_k\}_{k=1}^{N^2} \) can be represent as:  \begin{equation}\label{POVM_WHOLE_qubit} \begin{aligned}        E^{{\rm tot}}_ k =\vec{v}^{(1)}_{k^{(1)}}\otimes\vec{v}^{(2)}_{k^{(2)}}, \end{aligned} \end{equation} where, for $n=1,2$, \begin{equation}     \vec{v}^{(n)}_{k^{(n)}}=\frac{2}{N} \begin{pmatrix} 1  \\ \vec{r}^{(n)}_{k^{(n)}}   \end{pmatrix}. \end{equation}
Furthermore, we can derive the following representation for \(\Phi_E^\dagger\):
\begin{equation}
\Phi_E^\dagger = \frac{2}{N} \begin{pmatrix}
1 & 1 & \cdots & 1 \\
\vec{r}_1 & \vec{r}_2 & \cdots & \vec{r}_N
\end{pmatrix}.
\end{equation}
According to Eq. (\ref{8}), we obtain the least-square estimator operator:
\begin{equation}
\chi_{\rm LS} = \begin{pmatrix}
1 & 1 & \cdots & 1 \\
W^{-1} \vec{r}_1 & W^{-1} \vec{r}_2 & \cdots & W^{-1} \vec{r}_N
\end{pmatrix},
\end{equation}
where \(W\) is defined as:
\begin{equation}
W = \frac{1}{N} \sum_{k=1}^N \vec{r}_k \vec{r}_k^T.
\end{equation}
Notably, the columns of \(\chi_{\rm LS}\) correspond exactly to \(\hat{\rho}_k\):
\begin{equation}
\hat{\rho}_k = 
\begin{pmatrix}
1 & \\
W^{-1} \vec{r}_k 
\end{pmatrix}.
\end{equation}
Hence, the shadow of Choi state  can be represent as:
\begin{equation}
    \begin{aligned}
        \frac{\hat{\eta}_k}{d} &= \hat{\rho}^{(1)}_{k^{(1)}}\otimes \hat{\rho}^{(2)}_{k^{(2)}}
        \\& = \begin{pmatrix}1 &\\{W^{(1)}}^{-1} \vec{r}^{(1)}_{k^{(1)} }\end{pmatrix} \otimes \begin{pmatrix}1 &\\{W^{(2)}}^{-1} \vec{r}^{(2)}_{k^{(2)}} \end{pmatrix},
    \end{aligned}
\end{equation}
where, for $n=1,2$,
\begin{equation}
        W^{(n)} = \frac{1}{N} \sum_{k^{(n)}=1}^N \vec{r}_{k^{(n)}}^{(n)} [\vec{r}^{(n)}_{k^{(n)}}]^T.     
\end{equation}

 Notice that we always assume  the POVM has uniform trace, which is generally not a limiting assumption. In fact, if we start with a POVM that has effects with non-uniform traces, we can effectively transform it into one with uniform traces by splitting each effect into a suitable number of identical smaller effects. (For details, see  Appendix \ref{appendix splitting}.)

\begin{algorithm}[H]
\caption{Simulated Annealing with Hastings-Metropolis Step for POVM-SPT}
\begin{algorithmic}[1]
\STATE \textbf{Input:} Initial POVM in Bloch representation \( \mathbf{v} = \{\vec{v}_k\}_{k=1}^N =\{\frac{2}{N} \begin{pmatrix} 1  \\ \vec{r}_k  \end{pmatrix} \}_{k=1}^N\) , initial temperature $T_0$, minimum temperature $T_{\text{min}}$, cooling rate $\gamma$
\STATE \textbf{Output:} The optimal POVM $\mathbf{v_{\text{final}}}$ and  its corresponding  $\kappa^2_{\mathbf{v_{\text{final}}}}$
\WHILE{$T > T_{\text{min}}$}
    \FOR{$i = 1$ \text{ to } $n_{\text{steps}}$}
        \STATE \textbf{Select two random effects} $k_1, k_2 \in \{1, 2, \dots, N\}$
        \STATE Extract 3D components: $(x_1, y_1, z_1)^T \gets \vec{r}_{k_1}$, $(x_2, y_2, z_2)^T \gets \vec{r}_{k_2}$
        \STATE Generate Gaussian noise $\vec{\xi} = (\xi_1, \xi_2, \xi_3) \sim \mathcal{N}(0, \sqrt{T})$
        \STATE Calculate new effects:
        \STATE \quad $\vec{v}_{k_1}' \gets \frac{2}{N}(1, x_1 + \xi_1, y_1 + \xi_2, z_1 + \xi_3)^T$
        \STATE \quad $\vec{v}_{k_2}' \gets \frac{2}{N}(1, x_2 - \xi_1, y_2 - \xi_2, z_2 - \xi_3)^T$
        \STATE Check validity of the new vectors:
        \IF{either $\vec{v}_{k_1}'$ or $\vec{v}_{k_2}'$ is invalid: $\sum_{k=1}^N \vec{v}_k \ne  \mathbb{I}$}
            \STATE Reject new state: continue to next step
        \ENDIF
        \STATE Generate new POVM: \\ $\mathbf{v_{new}} = \{\vec{v}_{1}, \cdots,\vec{v}_{k_1}',\cdots, \vec{v}_{k_2}', \cdots ,\vec{v}_{N}  \}$
        \STATE \textbf{Calculate the energy difference:}
        \STATE \quad $Energy \gets \text{energy\_function}(\mathbf{v})=\kappa_{\mathbf{v}}^2$
        \STATE \quad $Energy' \gets \text{energy\_function}(\mathbf{v_{\text{new}}})=\kappa_{\mathbf{v_{new}}}^2$
        \STATE \quad $\Delta  \gets Energy' - Energy$
        \STATE \textbf{Acceptance Criterion:}
        \IF{$\Delta  < 0$ OR $\text{rand()} < \exp(-\Delta  / T)$}
            \STATE Accept new state: $\mathbf{v} \gets \mathbf{v_{\text{new}}}$
        \ENDIF
    \ENDFOR
    \STATE Cool down the temperature: $T \gets \gamma \cdot T$
   \IF{$T < 10^{-8}$}
       \STATE Break \COMMENT{End simulation if temperature is too low}
    \ENDIF
\ENDWHILE
\STATE \textbf{Return:}  $\mathbf{v_{\text{final}}}$ and $\kappa^2_{\mathbf{v_{\text{final}}}}$
\end{algorithmic}
\end{algorithm}

 The identification of optimal POVM $E^{\rm tot^*}$ can be formulated as a simulated annealing optimization problem where the system's configuration space corresponds to POVM parameters and the energy functional is defined by: \begin{equation}\label{kappa_min} \begin{aligned}   & \kappa^2_{E^{\rm tot} }(d \cdot \rho_l^T \otimes X_j) \\&=\max \{ \| d \cdot \rho_l^T \otimes X_j \|^2_{E^{\rm tot}} : d \cdot \rho_l^T \otimes X_j \in \mathcal{X} \} \\&=4 \times \max \{ \|  \rho_l^T  \|^2_{E^{(1)}} : 1\leq l \leq G \}  \\& \quad \times \max \{ \| X_j \|^2_{E^{(2)}} : 1 \leq j \leq H\} \\&=4 \kappa^2_{E^{(1)}}( \rho_l^T) \kappa^2_{E^{(2)}}( X_j). \end{aligned} \end{equation}  This energy functional decomposes into separable components for the system and ancilla, enabling independent optimization of $\kappa^2_{E^{(1)}}(\rho_l^T)$ and $\kappa^2_{E^{(2)}}(X_j)$. Through dimensional scaling and tensor product composition, the optimal POVM on the Choi state's composite system is constructed as specified in Eq. \eqref{POVM_WHOLE_qubit}.

The simulated annealing process initiates with high-temperature thermal fluctuations to escape local minima, employing Hastings-Metropolis dynamics \cite{reference1} to explore the constrained POVM space. At each thermalization stage with temperature $T$, random perturbations are generated through anisotropic Gaussian vectors $\vec{\xi}$
with standard deviation $\sqrt{T}$. These perturbations are applied symmetrically to pairs of randomly selected POVM effects $E_{k_1}$ and $E_{k_2}$, updating them as $E_{k_1} + \vec{\xi}$
and $E_{k_2} - \vec{\xi}$ while rigorously maintaining the identity constraint $\sum_{k=1}^N E_k = \mathbb{I}$. The energy difference $\Delta$ between new and old configurations determines acceptance probability through the Boltzmann factor $\exp(-\Delta/T)$, enabling both downhill moves and controlled uphill transitions. The temperature schedule follows exponential decay $T_{n+1} = 0.95T_n$, gradually reducing thermal noise to refine the solution landscape. This cooling protocol continues until reaching the terminal condition $T < 10^{-8}$, ensuring convergence to the global minimum corresponding to the optimal POVM configuration.

Without loss of generality, we consider the case where \( \{\rho_l\}_{l=1}^G \) is a set of single-qubit pure states. Although simple, this case provides a foundation for understanding multi-qubit scenarios.  For simplicity, we assume the POVM acting on both the system and ancilla has $N$  effects.  We analyze three cases: $N  =  4, 6$ and $8$. For each \( N \), our algorithm determines the optimal POVM \( E^{{\rm tot}^*} \) and computes the corresponding squared shadow norm \( \kappa^2_{E^{\mathrm{tot^*}}} \).
The observables are random projections distributed according to the Haar measure on the Bloch sphere, with the number of observables equal to the number of known states.
As shown in Fig. \ref{f3},  \( \kappa^2_{E^{\mathrm{tot^*}}} \) for \( N = 4 \) is consistently larger than for \( N = 6 \) and \( 8 \). Moreover, as the number of observables increases, \( \kappa^2_{E^{\mathrm{tot^*}}} \) for  \( N = 4 \) generally increases, while for \( N = 6 \) and \(  8 \), it converges to approximately $9$. Beyond a certain point, \( \kappa^2_{E^{\mathrm{tot^*}}} \) for  \( N = 6 \) and \(  8 \) become nearly indistinguishable, indicating that their values are almost equal in the limit of a large number of observables.

\begin{figure}[t]
  \centering
  \includegraphics[width=\linewidth]{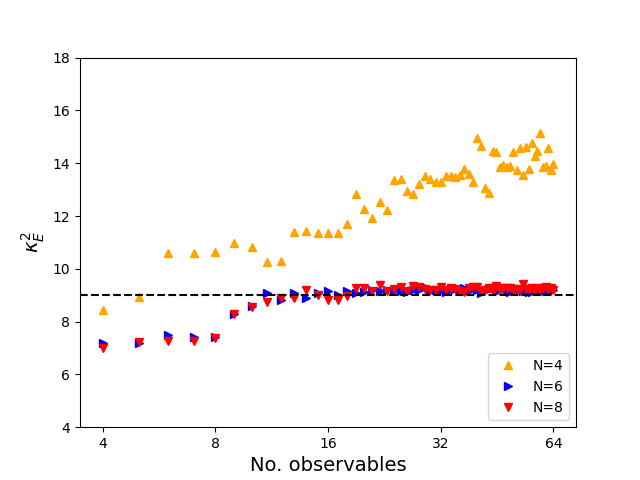}\\
  \caption{Dependence of the optimal squared shadow norm \( \kappa^2_{E^{\mathrm{tot^*}}} \) on the number of Haar-random projective observables for POVMs with \( N = 4 \), \( 6 \) and \( 8 \) effects. The number of observables is equal to the number of known states.}\label{f3}
\end{figure}

\begin{theorem}\label{theorem 3}
If the targeted observables \( \{X_j\}_{j=1}^H \) are all the projections on arbitrary single-qibit pure states and the known states \( \{\rho_l\}_{l=1}^G \) are all single-qubit pure states, when $H\to \infty$ or $G \to \infty  $, then $\kappa^2_{E^{\rm tot}} \geq 9$ and the optimal measurement is the tensor product of two octahedron measurements. (Proof:  Appendix \ref{appendix th3}.) 
\end{theorem}
\end{subsection}

The maximum square shadow norm $\kappa^2_{P^{\mathrm{tot^*}}}$ of traditional SPT exhibits four distinct cases corresponding to different measurement scenarios:
\begin{itemize}
    \item Scenario 1: $P^{\rm tot} = \mathcal{U}_{\rm P} \otimes \mathcal{U}_{\rm P}$
    \item Scenario 2: $P^{\rm tot} = \mathcal{U}_{\rm P} \otimes \mathcal{U}_{\rm C}$
    \item Scenario 3: $P^{\rm tot} = \mathcal{U}_{\rm C} \otimes \mathcal{U}_{\rm P}$
    \item Scenario 4: $P^{\rm tot} = \mathcal{U}_{\rm C} \otimes \mathcal{U}_{\rm C}$
\end{itemize}
where $\mathcal{U}_{\rm P}$ and $\mathcal{U}_{\rm C}$ denote PUM and CUM respectively. Under identical conditions to Fig.~\ref{f3}, we numerically obtain $\kappa^2_{P^{\mathrm{tot^*}}}$ values of 64, 224, 224, and 784 for scenarios 1--4 respectively. Taking the minimal value 64 as the representative $\kappa^2_{P^{\mathrm{tot^*}}}$ for traditional SPT, we observe $\kappa^2_{P^{\mathrm{tot^*}}} \approx 7 \times \kappa^2_{E^{\mathrm{tot^*}}}$, demonstrating POVM-SPT's superior performance in reducing sample complexity. This advancement substantially decreases resource requirements and facilitates practical implementation of QPT tasks.

\begin{subsection}{$n$-qubit systems}

POVM-SPT is specifically designed for \(n\)-qubit known states where the corresponding Choi states are \(2n\)-qubit states.   The protocol employs a set of POVMs  $\{ E^{(1)},  \ldots, E^{(n)}, \ldots, E^{(2n)} \}$,
where each POVM acting on the $i$-th qubit defined as \(E^{(i)} = \{ E^{(i)}_{k^{(i)}} \}_{k^{(i)}=1}^{N_i}\).
The global POVM $E^{\text{tot}}$ on the $2n$-qubit system is defined through outcome strings 
$k = \{ k^{(1)}, \ldots, k^{(2n)} \}$. In the Bloch representation, each effect decomposes as:
\begin{equation}\label{POVM_WHOLE} \begin{aligned}        E^{{\rm tot}}_ k =\vec{v}^{(1)}_{k^{(1)}}\otimes \cdots\otimes \vec{v}^{(2n)}_{k^{(2n)}}, \end{aligned} \end{equation} 
where, for $m=1,\cdots,2n$,
\begin{equation}     \vec{v}^{(m)}_{k^{(m)}}=\frac{2}{N} \begin{pmatrix} 1  \\ \vec{r}^{(m)}_{k^{(m)}}   \end{pmatrix} .\end{equation}
The shadow of Choi state can be represent as: \begin{equation}     \begin{aligned}         \frac{\hat{\eta}_k}{d} & = \begin{pmatrix}1 &\\{W^{(1)}}^{-1} \vec{r}^{(1)}_{k^{(1)} }\end{pmatrix}           \quad\otimes \cdots\otimes          \begin{pmatrix}1 &\\{W^{(1)}}^{-1} \vec{r}^{(2n)}_{k^{(2n)} }\end{pmatrix}     \end{aligned}, \end{equation}
where, for $m=1,\cdots,2n$,
\begin{equation}         W^{(m)} = \frac{1}{N} \sum_{k^{(m)}=1}^N \vec{r}_{k^{(m)}}^{(m)} [\vec{r}^{(m)}_{k^{(m)}}]^T .\end{equation}

For an $n$-qubit system, while observables $X$ are conventionally constructed through tensor product decompositions $X = \bigotimes_{i=1}^n X^{(i)}$,
the density matrix $\rho$ generally resists such complete separability. This fundamental distinction becomes crucial when analyzing the ``new observables" $d\cdot\rho^T \otimes X$---the separability characteristics of $\rho$ directly determine the physical interpretability of these composite observables.
Therefore, it is essential to discuss the separability of \(\rho\).

If $\rho$ is all separable, i.e. $\rho = \rho^{(1)} \otimes \cdots \otimes \rho^{(n)}$, we have:
\begin{equation}
    \begin{aligned}
        & \quad \,\, \| d \cdot \rho^T \otimes X \|^2_{E^{\rm tot}}
        \\& =4^n \cdot \| \rho^{(1)} \|^2_{E^{(1)}}  \cdots \| \rho^{(n)}\|^2_{E^{(n)}}
        \\& \quad \cdot \| X^{(n+1)} \|^2_{E^{(n+1)}} \cdots \| X^{(2n)} \|^2_{E^{(2n)}}.
    \end{aligned}
\end{equation}
If $\rho$ is an $n$-body genuinely entangled state, we have:
\begin{equation}\label{n-gen-ent}
    \begin{aligned}
        & \quad \,\, \| d \cdot \rho^T \otimes X \|^2_{E^{\rm tot}}
        \\& =4^n  \| \rho \|^2_{E^{\rm tot'}} 
        \| X^{(n+1)} \|^2_{E^{(n+1)}} 
         \cdots \| X^{(2n)} \|^2_{E^{(2n)}}
        \\& =4^n \lambda_{\text{max}} \{ \sum_{k^{(1)},\cdots,k^{(n)}=1}^{N} \text{Tr}[(\hat{\rho}^{(1)}_{k^{(1)}}  \otimes \cdots
        \otimes \hat{\rho}^{(n)}_{k^{(n)}})\rho^T]^2     \\& (E^{(1)}_{k^{(1)}} \otimes \cdots \otimes E^{(n)}_{k^{(n)}}) \} 
        \| X^{(n+1)} \|^2_{E^{(n+1)}}  \cdots \| X^{(2n)} \|^2_{E^{(2n)}}.
    \end{aligned}
\end{equation}

In addition to the two extreme cases of being completely separable and completely inseparable, there exists a more general scenario known as \(k\)-separability \cite{guojie}. We denote a state \(\rho\) as \(\rho_k\) if it is a \(k\)-body genuinely entangled state. Consequently, for an \(n\)-body state, there can be various configurations of separability. For instance, consider the case where \(\rho = \rho^{(1)} \otimes \rho_{n-2} \otimes \rho^{(n)}\). In this situation, we first need to decompose the \((n-2)\)-body genuinely entangled state \(\rho_{n-2}\) according to Eq. (\ref{n-gen-ent}). Following this, we can multiply it by the squared shadow norm of the remaining two separable bodies to obtain the squared shadow norm of the overall state \(\rho\):
\begin{equation}
    \begin{aligned}
        & \quad \,\, \| d \cdot \rho^T \otimes X \|^2_{E^{\rm tot}}
        \\& =4^n \cdot \| \rho^{(1)}\|^2_{E^{(1)}} \| \rho_{n-2} \|^2_{E^{\rm tot'}} \| \rho^{(n)}\|^2_{E^{(n)}} 
        \\& \quad \cdot \| X^{(n+1)} \|^2_{E^{(n+1)}}  \cdots \| X^{(2n)} \|^2_{E^{(2n)}}
        \\& =4^n \| \rho^{(1)}\|^2_{E^{(1)}} \lambda_{\text{max}} \{ \sum_{k^{(2)},\cdots,k^{(n-2)}=1}^{N-2} \text{Tr}[(\hat{\rho}^{(2)}_{k^{(2)}}   \otimes
        \\&\quad \cdots \otimes \hat{\rho}^{(n-1)}_{k^{(n-1)}})\rho_{n-2}^T]^2  (E^{(2)}_{k^{(2)}} \otimes \cdots \otimes E^{(n-1)}_{k^{(n-1)}}) \} 
        \\& \quad \cdot \| \rho^{(n)}\|^2_{E^{(n)}} \| X^{(n+1)} \|^2_{E^{(n+1)}} 
        \quad \cdots \| X^{(2n)} \|^2_{E^{(2n)}}
    \end{aligned}
\end{equation}

Therefore, for an $n$-qubit  state, we can always convert it into find optimal local POVM on each qubit system. 
For separable qubits, we directly compute the product of the optimal square shadow norms for each qubit. For the entangled parts, we can calculate the optimal square shadow norm for the entire entangled system. The measurements are performed as a tensor product based on the number of entangled qubits, ultimately yielding the corresponding optimal POVM $E^{{\rm tot}^*}$. Finally, by multiplying the results from both parts and scaling by the dimension, we obtain the optimal square shadow norm for the $2n$-qubit system.

\begin{figure}[t]
  \centering
  \includegraphics[width=\linewidth]{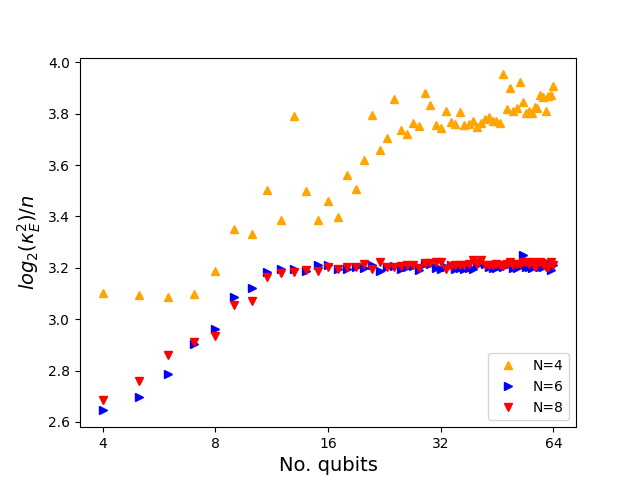}\\
  \caption{The relationship between the optimal  logarithmic squared shadow norm $\log_2(\kappa^2_{E^{\mathrm{tot^*}}})/n$
  and the number of qubits $n$ for three distinct types of POVMs, specifically in the context where all $n$-qubit quantum states are restricted to separable pure states.}\label{f4}
\end{figure}

Now, we consider a system of up to 64 qubits, assuming the states \( \{\rho_l\}_{l=1}^G \) are separable, such that \( \rho^{(1)} = \rho^{(2)} = \cdots = \rho^{(64)} \) are all pure qubit states. The observables \( X^{(1)} = X^{(2)} = \cdots = X^{(64)} \) are random projections distributed according to the Haar measure on the Bloch sphere. We calculate \( \frac{\log_2(\kappa^2_{E^{\mathrm{tot^*}}})}{n} \), where the number of observables equals the number of qubits in each calculation. To simplify the analysis without loss of generality, we assume \( E^{(1)} = \cdots = E^{(128)} \), with each \( E^{(i)} \) having \( N \) effects.
In Fig. \ref{f4}, we discuss three different cases: \( N = 4, 6 \) and $8$. Despite the increasing number of qubits, it is clear that \( \frac{\log_2(\kappa^2_{E^{\mathrm{tot^*}}})}{n} \) for \( N = 4 \) is always greater than that for \( N = 6 \) and \(  8 \). As the number of qubits increases, \( \frac{\log_2(\kappa^2_{E^{\mathrm{tot^*}}})}{n} \) for \( N = 4 \) roughly exhibits an increasing trend. However, the values of \( \frac{\log_2(\kappa^2_{E^{\mathrm{tot^*}}})}{n} \) for both \( N = 6 \) and \(  8 \) gradually converge to approximately 3.2 as the number of qubits increases, with  \( \frac{\log_2(\kappa^2_{E^{\mathrm{tot^*}}})}{n} \) for these two cases becoming nearly equal.

Under identical conditions to Fig.~\ref{f4}, our calculations reveal that the traditional SPT achieves a maximum square shadow norm of $\kappa^2_{P^{\mathrm{tot^*}}} = 2^{384}$, which corresponds to approximately $2^{180}$ times larger than $\kappa^2_{E^{\mathrm{tot^*}}}$ of POVM-SPT. This dramatic reduction demonstrates that POVM-SPT's advantage in sample complexity reduction becomes increasingly substantial with growing qubit numbers. These results establish crucial theoretical foundations for practical implementations of large-scale quantum computing and quantum networks.

\end{subsection}

\section{Conclusions and discussions}\label{sec5}
In this work, we present a generalized framework for SPT that utilizes POVMs combined with least-square estimators for reducing the complexity of SPT.
Unlike conventional SPT methods relying on randomized unitary operators and projective measurements, our approach enables convex optimization over POVMs to minimize the shadow norm, thereby achieving a provable reduction in sample complexity. 

Our numerical experiments demonstrate that POVM-SPT significantly outperforms standard SPT in shadow norm reduction. Specifically, we observe an approximately 7-fold reduction for single-qubit input states and a remarkable improvement of approximately $2^{180}$ for 64-qubit input states.
The key advantages of POVM-SPT include:  1) optimality: significantly reduce sample complexity. 2) generality: POVMs subsume projective measurements, allowing broader applicability; 3) practicality: eliminating the need for random unitaries simplifies experimental implementation;
Future research may investigate noise resilience in POVM optimization by analyzing practical noise effects like depolarization and readout errors. Another direction could extend POVM-SPT frameworks to quantify nonlinear properties such as entanglement entropy and state fidelity. Researchers might also adapt POVM-SPT for characterizing quantum operations within gate sequences, advancing gate set tomography methods. 

\section*{\textbf{ACKNOWLEDGMENTS}}
This work was supported by the National Natural Science Foundation of China (No. 12271394).

\appendix
\section{Mathematical characterization of classical shadows under symmetric measurements}\label{appendix A}
 A (finite) set of \( N \) rank-one projectors \( \{ |v_k\rangle \langle v_k| \}_{k=1}^N \) is termed a (complex-projective) two-design if it satisfies the condition:
\begin{equation}
    \frac{1}{N} \sum_{k=1}^N |v_k\rangle \langle v_k|^2 = \left( \begin{array}{c} d+1 \\ 2 \end{array} \right)^{-1} P_{\text{Sym}^{(2)}},
\end{equation}
where \( P_{\text{Sym}}(2) = \frac{1}{2} ( \mathbb{I} + \mathbb{F}) \). Here, \( \mathbb{F} \) represents the flip operator, defined by \( \mathbb{F} |x\rangle |y\rangle = |y\rangle |x\rangle \) for all \( |x\rangle, |y\rangle \in \mathbb{C}^d \).
Each 2-design is proportional to a symmetric POVM \( E = \{ \frac{d}{N} |v_k\rangle \langle v_k| \}_{k=1}^{N} \) \cite{ref2}.  When viewed as maps \( \Phi_E : \mathbb{H}_d \to \mathbb{R}^N \), these POVMs satisfy the relation \cite{ref2}:
\begin{equation}
\begin{aligned}
      \Phi_E^\dagger \Phi_E(O) &= \frac{d^2}{N^2} \sum_{k=1}^N \langle v_k | O | v_k \rangle |v_k\rangle \langle v_k| 
      \\& = \frac{d}{(d+1)N} [ O + \text{Tr}(O) \mathbb{I} ]
\end{aligned}
\end{equation}
for any operator \( O \) on \( \mathbb{H}_d \).
Therefore, we get:
\begin{equation}\label{8d}
\begin{aligned}
  \hat{\rho}_{k}&=(\Phi_E^\dagger \Phi_E)^{-1}(E_k) \\&= \frac{N}{d} [ (d + 1)E_k - \text{Tr}(E_k)\mathbb{I} ]
  \\&=\frac{N(d+1)}{d}E_k-\mathbb{I}.
\end{aligned}
\end{equation}

The first example we consider is the tetrahedral POVM \(M_{\text{tetra}} = \{ M_a = \frac{1}{4} \left( I + \mathbf{s}_a \cdot \boldsymbol{\sigma} \right) \}_{a \in \{0, 1, 2, 3\}}\)
, where
$
\mathbf{s}_0 = (0, 0, 1), \quad
\mathbf{s}_1 = \left( \frac{2\sqrt{2}}{3}, 0, -\frac{1}{3} \right), \quad
\mathbf{s}_2 = \left( -\frac{\sqrt{2}}{3}, \frac{2}{3}, -\frac{1}{3} \right), \quad
\mathbf{s}_3 = \left( -\frac{\sqrt{2}}{3}, -\frac{2}{3}, -\frac{1}{3} \right).
$
As the tetrahedron formed by these vectors is regular, it exemplifies a symmetric POVM. When \(N=4\) and \(d=2\), we can apply Eq. (\ref{8d}) to derive:
\begin{equation}
\hat{\rho}_{k}=6E_k-\mathbb{I}.
\end{equation}

The second example is the Pauli-6 POVM, which has effects:
$
E_0 = \frac{1}{3} \ket{0}\bra{0}, \quad E_1 = \frac{1}{3} \ket{1}\bra{1}, \quad E_2 = \frac{1}{3} \ket{+}\bra{+}, \quad E_3 = \frac{1}{3} \ket{-}\bra{-}, \quad E_4 = \frac{1}{3} \ket{r}\bra{r}, \quad E_5 = \frac{1}{3} \ket{l}\bra{l},
$
where \( \ket{0} \) and \( \ket{1} \) are the eigenstates of the Pauli operator \( \sigma_z \), \( \ket{+} \) and \( \ket{-} \) are the eigenstates of \( \sigma_x \), and \( \ket{r} \) and \( \ket{l} \) correspond to the eigenstates of \( \sigma_y \). Since the effects of this POVM form an octahedron on the Bloch sphere, it is also referred to as the octahedron measurement. For this POVM, with \(N=6\) and \(d=2\), we again utilize Eq. (\ref{8d}) to obtain:
\begin{equation}
\hat{\rho}_{k}=9E_k-\mathbb{I}.
\end{equation}

\section{The proof of Lemma \ref{lemma 1}}\label{appendix lem}
We provide a proof of Lemma  \ref{lemma 1} in this section.

\begin{proof}
By virtue of the Choi isomorphism, we have the relation $\text{Tr}[\frac{\eta}{d} (d \cdot \rho^T \otimes X)] = \operatorname{Tr}[\mathcal{E}(\rho) X]$. This allows us to view $d \cdot \rho^T \otimes X$ as an ``observable". For the normalized Choi state $\frac{\eta}{d}$, we can express the estimator as $\hat{x}_{k} = \text{Tr}[\frac{\hat{\eta}_k}{d} (d \cdot \rho^T \otimes X)] = \text{Tr}[\hat{\eta}_k (\rho^T \otimes X)]$. Utilizing results from Eq. (\ref{14}) and (\ref{15}), we can derive the inequality presented in Eq. (\ref{le1}):

\begin{equation}\label{le2}
\begin{aligned}
\text{Var}(\hat{x}_k) &= \sum_{k=1}^{N^2} \text{Tr}[\frac{\hat{\eta}_k}{d} (d \cdot \rho^T \otimes X)]^2 \text{Tr}(\frac{\eta}{d} E_k^{\text{tot}})- \langle X \rangle^2
\\& \leq \sum_{k=1}^{N^2} \text{Tr}[\hat{\eta}_k (\rho^T \otimes X)]^2 \text{Tr}(\frac{\eta}{d}E_k^{\text{tot}})
\\& =\sum_{k=1}^{N^2} \text{Tr}\left\{ \text{Tr}[\hat{\eta}_k (\rho^T \otimes X)]^2 \cdot \frac{\eta}{d}  E_k^{\text{tot}} \right\}
\\&=  \text{Tr} \left\{\frac{\eta}{d} \sum_{k=1}^{N^2} \text{Tr}[\hat{\eta}_k ( \rho^T \otimes X)]^2   E_k^{\text{tot}} \right\}
\\& \leq \max_{\frac{\eta}{d}} \text{Tr} \left\{\frac{\eta}{d} \sum_{k=1}^{N^2} \text{Tr}[\hat{\eta}_k( \rho^T \otimes X)]^2  E_k^{\text{tot}} \right\}
\\&  = \lambda_{\text{max}} \left\{ \sum_{k=1}^{N^2} \text{Tr}[\hat{\eta}_k (\rho^T \otimes X)]^2  E_k^{\text{tot}} \right\}.
\end{aligned}
\end{equation}
The final equality arises from the dual characterization of the spectral norm, which states that
\begin{equation}
\|A\|_\infty = \max_{\sigma:\, \text{state}} \text{Tr}(\sigma A)    
\end{equation}
 for any positive semidefinite matrix \( A \). This completes the proof of Lemma \ref{lemma 1}.

\end{proof}

\section{The proof of Theorem \ref{theorem 1}}\label{appendix th1}
We provide a proof of Theorem \ref{theorem 1} in this section.

\begin{proof}
The theorem is derived from combining the variance estimates presented in Lemma \ref{lemma 1} with a robust performance guarantee for the median of means estimation \cite{6}. Considering a random variable \( Y \) with variance \( \sigma^2 \), it can be shown that \( K \) independent sample means, each of size:
\begin{equation}\label{22}
M/K = \frac{34\sigma^2}{\epsilon^2},
\end{equation}
are sufficient to construct a median of means estimator \( \hat{\mu}(M/K, K) \) that satisfies:
\begin{equation}\label{23}
\Pr \left\{ |\hat{\mu}(M/K, K) - E[Y]| \geq \epsilon \right\}\leq 2e^{-K/2}
\end{equation}
for every \( \epsilon > 0 \).

We choose the parameters \( K \) and \( M/K \)  as Eq. (\ref{eq:K}) and Eq. (\ref{eq:M}) such that this general statement ensures:
\begin{equation}\label{24}
\Pr \left\{  | \hat{x}_j(\rho_l,M/K,K)- \operatorname{Tr}[\mathcal{E}(\rho_l) X_j] |  \geq \epsilon \right\} \leq \frac{\delta}{HG}
\end{equation}
for all $1\leq j\leq H, 1\leq l \leq G$.
Apply a union bound over all $HG$ failure probabilities to deduce:
\begin{equation}\label{25}
\Pr \left\{  | \hat{x}_j(\rho_l,M/K,K)- \operatorname{Tr}[\mathcal{E}(\rho_l) X_j] |  \leq \epsilon \right\} \geq 1-\delta
\end{equation}
for all $1\leq j\leq H, 1\leq l \leq G$. This completes the proof of Theorem \ref{theorem 1}.

\end{proof}

 \section{The proof of Theorem \ref{theorem 2}}\label{appendix th2} We provide a proof of Theorem \ref{theorem 2} in this section. 
 \begin{proof}     Fix an observable \(d \cdot \rho_l^T \otimes X_j\). Since any projection measurement belongs to the POVM set (see Fig. \ref{measurement}), we have:      \begin{equation}          \arg\min_{P^{\rm tot}} \|d \cdot \rho_l^T \otimes X_j \|^2_{P^{\rm tot}} \geq \arg\min_{E^{\rm tot}} \|d \cdot \rho_l^T \otimes X_j \|^2_{E^{\rm tot}}.     \end{equation}     Thus, we obtain:     \begin{equation}     \begin{aligned}         &\quad \kappa^2_{P^{\rm tot^*}} \\&= \max \left\{ \| d \cdot \rho_l^T \otimes X_j \|^2_{P^{\rm tot^*}} : d \cdot \rho_l^T \otimes X_j \in \mathcal{X} \right\} \\         &\geq \max \left\{ \| d \cdot \rho_l^T \otimes X_j \|^2_{E^{\rm tot^*}} : d \cdot \rho_l^T \otimes X_j \in \mathcal{X} \right\} \\         &= \kappa^2_{E^{\rm tot^*}}.     \end{aligned}     \end{equation} \end{proof}

\section{Transforming measurements with non-uniform traces into those with uniform traces}\label{appendix splitting}

Specifically, consider a measurement \( E \) where the traces for each effect are given by \( \alpha_k = \text{Tr}(E_k) \). These trace values can be approximated by rational numbers. By selecting a sufficiently small value \( \epsilon \), we can ensure that \( \frac{\alpha_k}{\epsilon} = N_k \) are all integers for each effect. 
We can then divide the effect \( E_k \) into \( N_k \) identical effects, represented as follows:
\begin{equation}
\left( \overbrace{\frac{E_k}{N_k}, \frac{E_k}{N_k}, \dots, \frac{E_k}{N_k}}^{N_k} \right).    
\end{equation}
This process leads to a new measurement with a total of \( N_{\text{tot}} = \sum_{k=1}^N N_k \) effects. 
The new measurement $E'= \{ E'_{k'} \}_{k'=1}^{N_{\text{tot}}}$ can be expressed as:
\begin{equation}
\begin{aligned}
    \left\{ \overbrace{\frac{E_1}{N_1},  \dots, \frac{E_1}{N_1}}^{N_1}, \cdots, \overbrace{\frac{E_k}{N_k}, \dots, \frac{E_k}{N_k}}^{N_k}, 
      \cdots, \overbrace{\frac{E_N}{N_N}, \frac{E_N}{N_N}}^{N_N} \right\},
\end{aligned}  
\end{equation}
which has uniform trace.

 \section{The proof of Theorem \ref{theorem 3}}\label{appendix th3} We provide a proof of Theorem \ref{theorem 3} in this section. 
 \begin{proof} For single qubit system, the dimension is $d=2$. Then, we have: \begin{equation} \begin{aligned} & \kappa^2_{E^{\rm tot} }(d \cdot \rho_l^T \otimes X_j) \\&=\max \{ \| d \cdot \rho_l^T \otimes X_j \|^2_{E^{\rm tot}} : d \cdot \rho_l^T \otimes X_j \in \mathcal{X} \} \\&=\max \{ d^2 \cdot \|\rho_l^T  \|^2_{E^{(1)}} \cdot \| X_j \|^2_{E^{(2)}}: 1\leq l \leq G,\,1 \leq j \leq H \} \\&=4 \times \max \{ \|  \rho_l^T  \|^2_{E^{(1)}} : 1\leq l \leq G \}  \\& \quad \times \max \{ \| X_j \|^2_{E^{(2)}} : 1 \leq j \leq H\} \\&=4 \kappa^2_{E^{(1)}}( \rho_l^T) \kappa^2_{E^{(2)}}( X_j) \end{aligned} \end{equation} According to Ref. \cite{0}, we know that if the targeted observables are all the projections on arbitrary pure states of the qubit, then, $\kappa^2_{E^{(i)}} \geq 3/2$ and the optimal measurement would be the octahedron measurement assuming equal trace of the effects. Therefore, for POVM-SPT, we get $\kappa^2_{E^{\rm tot}} \geq 9$. The optimal measurement is the tensor product of two octahedron measurements.
 
 \end{proof}

\section{The proof pf Eq. (\ref{lel-p}).}\label{appendix b}
Let $\mathbb{E}_{U \in \mathbf{U}} [ U]=U_0$ where $\mathbb{U}$ is an  ensemble of unitary operators, $\{|k\rangle\}_1^{N^2}$ is the computational basis on two-qubit system,  \( C = \Phi^\dagger \Phi \) is defined as main text, $P^{{\rm tot}}_k=U_0^{\dagger}|k\rangle \langle k|U_0 $. According the definition of shadow norm in \cite{ref3,3}, we get:
\begin{widetext}
\begin{equation}
    \begin{aligned}
        &\quad \,\|d \cdot \rho^T \otimes X \|^2_{P^{\rm tot}} 
        \\&= \max_{\frac{\eta}{d}}  \left\{ \mathbb{E}_{U \in \mathbf{U}} 
        [ \sum_{k=1}^{N^2} \langle k|U \frac{\eta}{d} U^{\dagger}|k\rangle
         \langle k|U C^{-1}(d\cdot\rho^T \otimes X) U^{\dagger}|k\rangle^2 ]\right\}
        \\&= \max_{\frac{\eta}{d}}  \left\{ \sum_{k=1}^{N^2} \langle k|U_0 \frac{\eta}{d} U_0^{\dagger}|k\rangle
         \langle k|U_0 C^{-1}(d\cdot \rho^T \otimes X) U_0^{\dagger}|k\rangle^2 \right\}
        \\& =\max_{\frac{\eta}{d}}  \left\{ \sum_{k=1}^{N^2} \Tr( \frac{\eta}{d} U_0^{\dagger}|k\rangle \langle k|U_0) 
         \Tr[ C^{-1}(d\cdot\rho^T \otimes X) U_0^{\dagger}|k\rangle \langle k|U_0]^2 \right\}
         \\&=\max_{\frac{\eta}{d}}  \left\{ \sum_{k=1}^{N^2} \Tr( \frac{\eta}{d} U_0^{\dagger}|k\rangle \langle k|U_0) 
         \Tr[ (d\cdot\rho^T \otimes X )C^{-1}( U_0^{\dagger}|k\rangle \langle k|U_0)]^2 \right\}
         \\& =\max_{\frac{\eta}{d}}  \left\{ \sum_{k=1}^{N^2} \Tr \{ \frac{\eta}{d}   \Tr[ d(\rho^T \otimes X) 
           C^{-1}( U_0^{\dagger}|k\rangle \langle k|U_0)]^2 U_0^{\dagger}|k\rangle \langle k|U_0 \} \right\}
           \\&=\max_{\frac{\eta}{d}}  \left\{ \sum_{k=1}^{N^2} \Tr \{ \frac{\eta}{d}   \Tr[ d\cdot(\rho^T \otimes X) \frac{\hat{\eta}_k}{d}]^2 U_0^{\dagger}|k\rangle \langle k|U_0 \} \right\}
          \\& =\max_{\frac{\eta}{d}}  \left\{  \Tr \{ \frac{\eta}{d} \sum_{k=1}^{N^2}  \Tr[(\rho^T \otimes X) \hat{\eta}_k]^2 U_0^{\dagger}|k\rangle \langle k|U_0 \} \right\}
        \\&= \lambda_{\text{max}} \left\{ \sum_{k=1}^{N^2} \text{Tr}[\hat{\eta}_k (\rho^T \otimes X)]^2  U_0^{\dagger}|k\rangle \langle k|U_0 \right\}
        \\&= \lambda_{\text{max}} \left\{ \sum_{k=1}^{N^2} \text{Tr}[\hat{\eta}_k (\rho^T \otimes X)]^2  P^{{\rm tot}}_k \right\}.
    \end{aligned}
\end{equation}
\end{widetext}
The fourth equality arises from $C^{-1}$ being self-adjoint:
    $\Tr[AC^{-1}(B)]=\Tr[C^{-1}(A)B]$ for any compatible-dimension matrices $A,B$.
The final equality arises from the dual characterization of the spectral norm, which states that
$\|A\|_\infty = \max_{\sigma:\, \text{state}} \text{Tr}(\sigma A) $
for any positive semidefinite matrix \( A \).

\end{document}